\documentclass[aps,pra,reprint,groupedaddress,amsmath,amssymb,floatfix]{revtex4-1}

\usepackage{graphicx}   
\usepackage{epstopdf}
\graphicspath{{img/}}

\usepackage{braket}

\newcommand{\superscript}[1]{\ensuremath{^{\textrm{#1}}}}

\begin{document}.


\title{Compressibility of an Ultracold Fermi Gas with Repulsive Interactions}



\author{Ye-Ryoung Lee$^1$, Myoung-Sun Heo$^1$, Jae-Hoon Choi$^1$, Tout T. Wang$^2$, Caleb A. Christensen$^1$, Timur M. Rvachov$^1$, and Wolfgang Ketterle$^1$}

\affiliation{$^1$MIT-Harvard Center for Ultracold Atoms, Research Laboratory of Electronics, Department of Physics, Massachusetts Institute of Technology, Cambridge, MA 02139, USA}
\affiliation{$^2$MIT-Harvard Center for Ultracold Atoms, Department of Physics, Harvard University, Cambridge, MA 02138, USA}





\begin{abstract}

Fermi gases with repulsive interactions are characterized by measuring their compressibility as a function of interaction strength.
The compressibility is obtained from in-trap density distributions monitored by phase contrast imaging. For interaction parameters $k_F
a > 0.25$ fast decay of the gas prevents the observation of equilibrium profiles.  For smaller interaction parameters, the results are
adequately described by first-order perturbation theory. A novel phase contrast imaging method compensates for dispersive distortions
of the images.

\end{abstract}


\maketitle


\section{\label{intro}Introduction}

Experiments with ultracold atoms explore many-body physics with strong interactions.  They have demonstrated long-predicted
phenomena like the BEC-BCS crossover \cite{varenna2008} and Lee-Huang-Yang corrections to the energy of
degenerate gases \cite{ShinLHY,ENS_BEC,ENS_science}. Experiments have also explored novel
quantum phases like fermions with unitarity limited interactions
\cite{varenna2008,Mark2012}, population imbalanced Fermi gases
\cite{imbalance_MIT, imbalance_JILA} and Hubbard models in optical
lattices \cite{hubbard,BlochRMP80}. More recently, they have been used to
provide precision tests of many-body theories \cite{Mark2012}. Usually,
interactions in ultracold gases are fully described by the
scattering length, which is a zero-range approximation greatly
simplifying the theoretical description. This approximation is
valid since the diluteness of the atomic gases implies a
particle separation much larger than the short range of the
van-der-Waals interactions. This almost exact characterization of
the interactions by a single parameter, the tunability of interaction strength, and precise experimental control
over cold atoms systems have
made them an ideal testbed for many-body quantum calculations.

A new level of quantitative comparison between theory and experiment was recently reached by careful measurements of density profiles from which the equation of state could be determined.  These techniques were first proposed by Chevy \cite{Chevy} and Bulgac
\cite{Bulgac} and implemented by Shin \cite{Yong_EoS}.  Further improvements \cite{Ho, ENS_nature, ENS_science, ENS_sus, tokyo,Mark2012}
resulted in impressive accuracy without  adjustable parameters. These results hinge on accurate measurements of the equilibrium atomic density distribution.  Since all cold atom systems are in a metastable phase, this requires a favorable ratio of lifetime to
equilibration time.

Long lifetimes and strong interactions were realized in Fermi gases with strong attractive interactions since the decay to lower-lying
molecular states is suppressed by the Pauli exclusion principle \cite{petrov2004}. This is different for repulsive interactions which are realized on the so-called
upper branch of a Feshbach resonance where decay is always possible into the so-called lower branch which consists of weakly bound
molecular states with binding energy $\hbar^2/m a^2$ with $a$ being the scattering length.

For bosons, the first beyond mean field correction, the so-called Lee-Huang-Yang term, could be observed, but corrections were
necessary to account for the non-equilibrium profile since the time to sweep to strong interactions was not long compared to
equilibration times and inverse loss rates \cite{ENS_BEC}.  Here we study fermions with repulsive interactions.  They have been the
focus of much recent work due to the prediction of a phase transition to a ferromagnetic state for sufficiently strong
interactions \cite{Mcdonald,stoner}.  Recent experimental \cite{FM,BEC2sus} and theoretical studies \cite{Pekker2011} addressed the competition with strong decay to the lower molecular branch.

As expected we find only a limited window for metastability where we can observe equilibrated clouds and characterize the repulsive
interactions by obtaining the compressibility from observed profiles. We observe the linear term in $k_Fa$ corresponding to mean field energy for
the first time in density profiles. $k_F$ is the Fermi wave vector. In contrast to a Bose-Einstein condensate, the mean field energy is smaller than the kinetic energy
and also competes with thermal energy, and is therefore much more difficult to observe.  The signal-to-noise ratio (and some heating)
prevented us from discerning the second-order interaction term which is the Lee-Huang-Yang correction for fermions.

Our work features one technical novelty, a novel implementation of
phase-contrast imaging to address dispersive distortions of the
cloud.  All studies mentioned above, with one exception
\cite{Yong_EoS}, were conducted using resonant absorption where
dispersion (an index of refraction different from one) is absent.
However, this severely limits the cloud size and number of atoms
to optical densities of a few.  Phase-contrast imaging has many
advantages. It can be applied to clouds with much larger optical
densities by adjusting the detuning.  Due to coherent forward
scattering, the heating effect per detected signal photon is
reduced by potentially a large number (which is equal to the
resonant optical density divided by four \cite{varenna1999}). This
can e.g. be used for repeated nondestructive imaging. However, for
precision studies of density profiles, small dispersive
distortions of the density profile cannot be
neglected.  Previous work including Ref. \cite{Yong_EoS} was not
sensitive to this effect. We have developed an experimental
technique to correct for dispersion.

\section{\label{ex}Experimental Setup}

A spin polarized Fermi gas of \superscript{6}Li in the $\ket{F=3/2; m_{F}=3/2}$ state is produced by sympathetic cooling with bosonic \superscript{23}Na
atoms in a magnetic trap as described in \cite{Zoran}. The \superscript{6}Li atoms are then loaded into a single-beam optical dipole trap and
transferred into the lowest hyperfine state $\ket{F=1/2; m_{F}=1/2}$ by a Landau-Zener radio-frequency (RF) sweep. Additional axial confinement is provided
by magnetic field curvature. An equal mixture of $\ket{1}$ and $\ket{2}$ spin states (corresponding to the $\ket{F=1/2; m_{F}=1/2}$ and
$\ket{F=1/2; m_{F}=-1/2}$ states at low magnetic field) is prepared by a RF sweep at 300 G, followed by 500 ms wait
time for decoherence and evaporative cooling in the optical trap. The Feshbach resonance at 834 G \cite{varenna2008} is used to tune the repulsive interactions between $\ket{1}$ and $\ket{2}$. We increase the magnetic field in 200 ms to 528 G, where the scattering length is zero and our Fermi gas is non-interacting. The final trap has a depth of 4.1 $\mu$K and frequencies
of $\omega_{x}=\omega_{y}$= 390 Hz and $\omega_{z}$= 34.7 Hz. The number of atoms per spin state is $8 \times 10^5$,
which corresponds to a Fermi temperature $T_F$ of 1.4 $\mu$K. The temperature of the atoms is 0.3 $T_F$ at this point.  For
loss rate measurements, the magnetic field is quickly ramped to the target field.  For compressibility  measurements, the field is
ramped up over 50 ms and held for 30 ms to ensure thermal equilibrium before imaging. The molecular fraction in the density profile is determined by dissociating molecules with a magnetic field jump to 20 G above the Feshbach resonance, and comparing with the atom number after jumping to 528 G where the cross section for imaging molecules vanishes (see Ref. \cite{dissociation}).

\section{Loss Rate}

\begin{figure}
\includegraphics[width=3.3in]{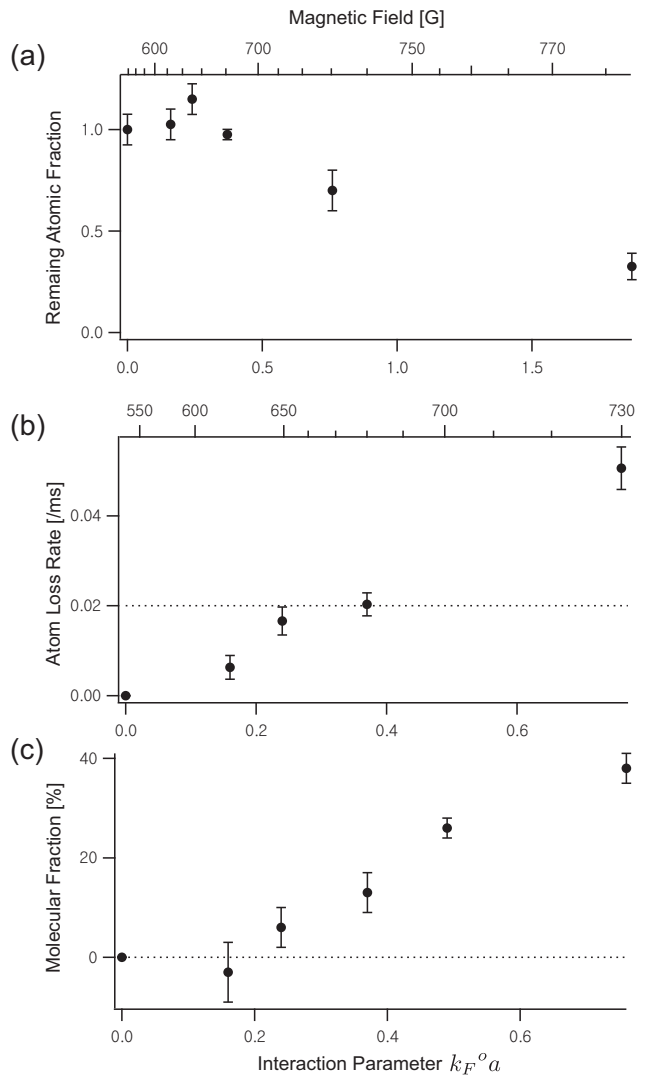}
\caption{Characterizing atomic loss for increasing repulsive interactions.
(a) Remaining fraction of atom number immediately after the fastest possible ramp to
the target field. (b) Atom loss rates at the target fields. Dashed line shows the
estimated maximum tolerable loss rate of 0.02 /ms. (c) Molecular fraction after the 50
ms ramp and 30 ms wait time, corresponding to when we measure equilibrium atomic density profiles.} \label{loss}
\end{figure}

To obtain thermodynamic parameters from atomic density profiles requires
equilibration.  One time scale for equilibration is set by the
longest trap period, which is 30 ms for the axial direction. Ref.
\cite{ENS_BEC} studied the distortions of profiles of bosonic
\superscript{7}Li due to non-adiabatic ramping of the scattering
length.  The authors found that ramping times longer than $\approx
6 \pi/ \omega$ (which is 90 ms for our parameters) led to only
small deviations from equilibrium profiles.  Assuming that losses
sharply increase during the last 5 ms of the ramp towards stronger
interactions and limiting tolerable losses to 10 \% leads to an
estimate of a maximum tolerable loss rate of about 0.02/ms. The
fastest relaxation time for excitations created during a
non-adiabatic ramp is $1/\omega$ (the damping time for a harmonic
oscillator at critical damping).  Allowing $\sim$ 10 $\%$ loss
during this time, leads to an identical estimate for the tolerable
loss rate of $\sim$ 0.02/ms.

We measured loss rate as a function of interaction strength ${k_F}^o a$.  Note that ${k_F}^o$ is the Fermi wave vector of the zero
temperature noninteracting gas calculated at the trap center using total atom number. The real $k_F$ is somewhat smaller because
repulsive interactions and non-zero temperature lower the density.

First, in Fig.~\ref{loss}(a) we measured the number of atoms right before and after the fastest possible ramp (limited to 7 ms by eddy currents) to the
target magnetic field. During the ramp to the target magnetic field of ${k_F}^o a$ $\approx$ 0.8, $\sim$ 35 $\%$ of the sample is lost.
Measuring the loss rate at higher ${k_F}^o a$ is difficult because most of the sample is lost even before reaching the target fields.
The loss rate was determined by  monitoring the drop in the number of atoms immediately after the field ramp. The results in Fig.~\ref{loss}(b) show that the
measured loss rate reaches the maximum tolerable value of 0.02/ms at ${k_F}^o a$ $\approx$ 0.35, limiting our measurements of
equilibrium density profiles to smaller values of ${k_F}^o a$.  Furthermore, at the same values of ${k_F}^o a$, the
molecular fraction when we measure equilibrium density profiles exceeded 10 $\%$ (see Fig.~\ref{loss}(c)).  As we discuss in Sec.~\ref{sec:comp}  the presence of a molecular component affects the compressibility measurement.

\section{Compressibility Measurement}\label{sec:comp}

A system is characterized by its equation of state.  The equation of state can be expressed in different forms involving various
thermodynamic variables including density, energy, pressure, temperature and entropy.  For cold atom experiments, density, chemical
potential (through the trapping potential) and temperature are directly accessible to measurement.  In the weakly interacting regime, the interaction manifests itself as a perturbative term in the equation of state. Here we measure this perturbative interaction effect by measuring the derivative of
density with chemical potential, the isothermal compressibility of the gas.  We prepare the system at the lowest temperature, but due
to heating of the cloud by molecule formation, we have to apply a correction for the measured finite temperature.

Using the experimental procedure discussed above we prepared
equilibrated clouds at various magnetic fields and measured
line-of-sight integrated profiles of column density using in situ
phase-contrast imaging \cite{pc_shin}. The signal-to-noise was
improved by averaging the column density along equipotential lines
(which are ellipses for the anisotropic harmonic oscillator
potential). The averaging region was restricted to an axial sector
of $\pm 60^{\circ}$ to avoid corrections due to transverse
anharmonicities \cite{Yong_nature}. Three-dimensional density
profiles $n(r)$ were reconstructed by applying the inverse Abel
transformation to the column densities $\tilde{n}(r)$
\cite{pc_shin}.

The isothermal compressibility is obtained from a spatial derivative of $n(r)$, since in the local density approximation, the local
chemical potential is  $\mu=\mu_{o}- m \omega^{2} x^{2}/2$, where $\mu_{o}$ is the global chemical potential and $\omega$ the trap
frequency.  The compressibility is defined as
\begin{eqnarray}
\kappa = && \frac{1}{n^2}\frac{\partial n}{\partial \mu}.
\end{eqnarray}
We normalize the compressibility $\kappa$ by the compressibility of an ideal gas at the given density and zero
temperature $\kappa_{o}= \frac{1}{n^2} \frac{3m}{\hbar^2 {(6 \pi^2)}^{2/3}
n^{1/3}}$, and obtain the normalized compressibility,
\begin{eqnarray}\
\tilde{\kappa}=\frac{\kappa}{\kappa_o} = && \frac{\hbar^2 (6 \pi^2)^{2/3}}{2 m}\frac{\partial n^{2/3}}{\partial \mu}.
\end{eqnarray}
Here $n$ is density, and $m$ is the atomic mass. The normalized
compressibility is obtained as the slope in a plot of $n^{2/3}$
vs. $\mu$ (Fig. 2).  This plot is in essence the observed density
profile plotted with the central region to the left and the
spatial wings to the right.  Experimentally, we find the slope to
be constant over an extended range of the density profile.
Compressibilities were determined from fits to the slope in the
region of 90 to 50 $\%$ of the peak density.  The region near the
center of the cloud was excluded since the center is singular for
the inverse Abel transformation leading to excessive noise.  These
compressibilities should be regarded as average values over the
density range used in the fit. The uncertainty of fitting the
slope to a single profile was 4.5 \%. By averaging the profiles
obtained from 20 images, the uncertainty was improved to 1.3 \%.

The  normalized compressibility is a function of $T/T_F$ and $k_F
a$. At finite temperature and scattering length $a$, $T/T_F$ and
$k_F a$ change across a single density profile because $T_F$ and
$k_F$ depend on the local density.  However, this density dependence is
small near the center of the cloud.  Simulated density profiles
showed that the average compressibility determined in the way
described above agrees to within 0.6 $\%$ with the compressibility
at $T/T_F$ and $k_F  a$ at the density in the center of the
selected range.

\begin{figure}[!htbp]
\centering
\includegraphics[width=3.3in]{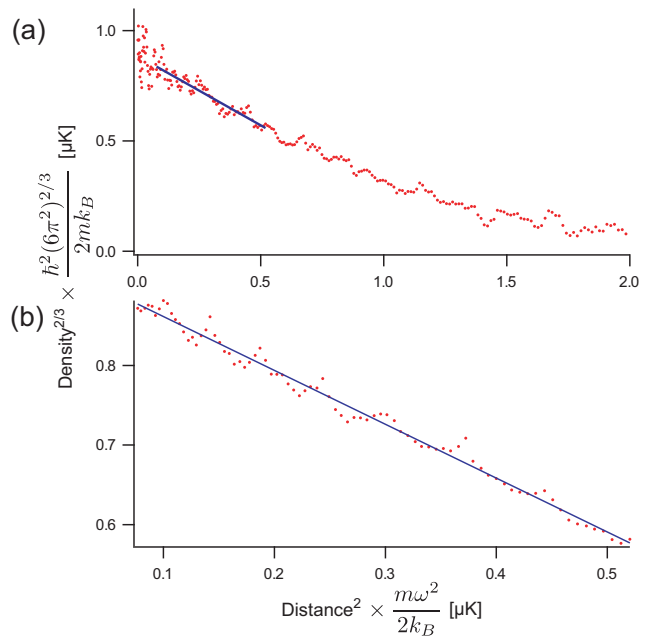}
\caption{Determination of the compressibility of repulsively interacting Fermi gases.  The compressibility is the slope of a graph
showing the density to the power 2/3 versus the square of the distance from the center.  (a) Single shot density profile.  (b) Close-up
of the region used for determining the compressibility for a plot showing the average of 20 density profiles.}
\label{comp}
\end{figure}

Compressibility decreases for stronger repulsion, but also for higher temperature.  To identify
the effect of repulsive interaction requires a careful consideration of finite temperature effects. First, the temperature of the cloud
had to be accurately determined.  This can be done without any special assumptions by fitting the wings of the cloud using a virial
expansion \cite{virial}, by thermometry with another co-trapped atom \cite{ENS_nature}, or for population imbalanced clouds by fitting
the wings of the majority component which is an ideal Fermi gas \cite{wing_fit}. Here we chose to determine temperature using a virial
expansion,
\begin{eqnarray}
p \frac{\lambda^3}{k_B T} = e^{\beta \mu} + b_2 e^{2 \beta \mu} + O(e^{3 \beta \mu}),
\end{eqnarray}
where $\lambda$=$\sqrt{\frac{2 \pi \hbar^2}{m k_B T}}$ is the thermal de Broglie wavelength, $b_2$ is the virial
coefficient, and $e^{\beta \mu}$ is the fugacity. The virial
coefficient for the Fermi gas with repulsive interaction is $b_2 =
- 2^{-5/2} - a/\lambda$ \cite{virial_coef}. Pressure $p$ was obtained
from the doubly-integrated density profiles \cite{Ho}. Temperature
was determined in the wings of the profile where $\beta \mu <
-0.5$. Here, the temperature measured with and without the
interaction term $-a/\lambda$ in $b_2$ differ by about
3$\%$. This suggests that higher-order corrections from interaction
term will be negligible.  Note that the virial expansion up to
second order is valid to within 1$\%$ for the  ideal gas at the density of the fitted wings.

The low temperature normalized isothermal compressibility of a non-interacting Fermi gas is given by the Sommerfeld expansion
\cite{sommerfeld},
\begin{eqnarray}
\tilde{\kappa}_{o,T} = && 1 - \frac{\pi^2}{12}\left(\frac{T}{T_F}\right)^2 + O\left[\left(\frac{T}{T_F}\right)^4\right],
\end{eqnarray}
where $T$ is temperature and $T_F$ is Fermi temperature.
To add the effects of interactions, it is useful to work with the inverse normalized compressibility,
\begin{eqnarray}
\frac{1}{\tilde{\kappa}} = && \frac{3}{2} \frac{2m}{\hbar^2 (6 \pi^2)^{2/3}} n^{1/3} \frac{\partial \mu (n, T, a)}{\partial n}.
\end{eqnarray}
This is a derivative of the chemical potential, which has the following expansion in temperature and scattering length,
\begin{eqnarray}
\mu (n, T, a) &&= E_F \left[1 - \frac{\pi^2}{12} \left(\frac{T}{T_F}\right)^2 + \frac{4}{3 \pi} k_F a \right. \nonumber\\
&&\left.+ \frac{4(11-2\ln2)}{15 \pi^2} (k_F a)^2\right] + C T^2 a^2,
\end{eqnarray}
where C is constant, independent of density \cite{second_order}. Therefore
the inverse normalized compressibility has additive correction terms for temperature and interactions up to the second order of the
interaction effect,
\begin{eqnarray}
\frac{1}{\tilde{\kappa}} = && \frac{1}{{\tilde{\kappa}}_{o,T}} + Y
(k_F a) \label{int_eq}.
\end{eqnarray}
This equation defines $Y (k_F a)$, the interaction correction to the inverse compressibility.  This term is the derivative of the
interaction term of the chemical potential.  In second order perturbation theory, one obtains  $Y (k_F a) = \frac{2}{\pi} k_{F}a +
\frac{8 (11-2 ln2)}{15 \pi^2} (k_{F}a)^{2}$.

\begin{figure}[tbp]
\includegraphics[width=3.3in]{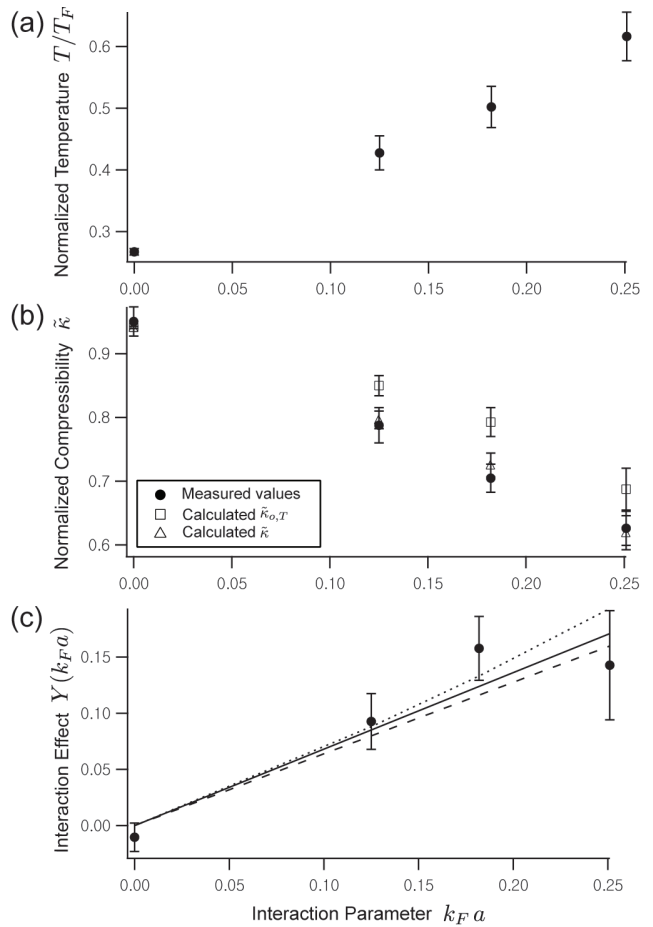}
\caption{Measured temperature, normalized compressibility, and
interaction effect on compressibility at various interaction
strengths. (a) The measured temperature as a function of
interaction strength. (b) Solid circles show the measured
normalized compressibility ($\tilde{\kappa}$) and open squares
show the calculated normalized compressibility at the measured
temperature without interaction (${\tilde{\kappa}}_{o,T}$). The
difference between the two indicates the effect of interaction.
Open triangles show the calculated normalized compressibility
using the second order perturbation theory, which is consistent
with our measured $\tilde{\kappa}$. (c) The measured interaction
correction to the inverse compressibility (solid circle) is
compared to a linear fit (solid line), and the first (dashed line)
and the second (dotted line) order perturbative result. }
\label{result}
\end{figure}

Fig.~\ref{result} shows the normalized compressibility, the temperature $T/T_F$, and the interaction correction to the inverse
compressibility.  The temperature increases with $k_F a$ due to the increase in three-body recombination where the binding energy of
the molecules ($\hbar^2/ma^2$) is transferred to the remaining atoms. The measured temperature is higher than that in previous
experiments on the repulsive side \cite{FM, BEC2sus}. This difference can be explained by a smaller $k_F a$ since the increase in
$T/T_F$ is approximately proportional to $1/k_F a$ \cite{BEC2sus}.

We perform a linear fit of the interaction effect $Y (k_F a)$ versus $k_F a$
(constrained to pass through the origin) and obtain
0.680$\pm$0.147 for the slope, in agreement with the perturbative
prediction of $\frac{2}{\pi}= 0.637$.  This is the first
observation of the mean-field term for repulsively interacting
fermions in a thermodynamic quantity. The repulsive interaction
has been seen as line shifts in RF spectroscopy experiments (which, in contrast
to many thermodynamic quantities, are measured independently of the kinetic
energy of the atoms) \cite{gupta, Jin}. In principle, it is
possible to obtain the mean-field term directly by fitting the
density profiles with an extra mean field term. In such fits, we
obtained clear evidence for such a term, but with low accuracy. It
appears that the averaging over profiles for determining the
compressibility(as in Fig.~\ref{comp}) is superior. Figure \ref{result} (c) shows the
predicted effect of the second-order term on $Y (k_F a)$. With some improvements in signal-to-noise ratio,
one should be able to observe this term which is analogous to the
Lee-Huang-Yang correction for bosons.

\section{Dispersive effect in phase-contrast imaging}

As mentioned in the introduction, phase-contrast imaging has
several advantages over resonant absorption imaging, and it has
been applied to many studies of cold Bose and Fermi gases
\cite{pc_andrew, pc_stamper, pc_shin}.  Absorption imaging is
usually done with absorptively dilute clouds, typically with 10 to
70 \% absorption (or optical densitites $OD<1$).  The standard
assumption has been that dispersive imaging is quantitative when
the phase shift $\phi$ across the cloud is less than $\pi/4$. The
normalized phase-contrast signal (for negligible absorption) is $3
- 2 \sqrt{2} \cos(\phi \pm \pi/4)$, which is equal to $1 \pm 2
\phi$ for small phase shifts \cite{varenna1999}.  The sign depends
on laser detuning and the sign of the phase shift imparted by the
phase plate.

Here, we apply phase-contrast imaging for rather precise quantitative studies of ultracold Fermi gases and found that even for small
phase shifts systematic dispersive distortions of the image cannot be neglected. Phase-contrast imaging relies on column density
dependent phase shifts.  However, if the object is not thin, but extended along the line of sight, some lensing will affect the images.
These distortions should vary inversely proportional to the probe light detuning and become negligible for far detuning.

\begin{figure}[!htbp]
\centering
\includegraphics[width=3in]{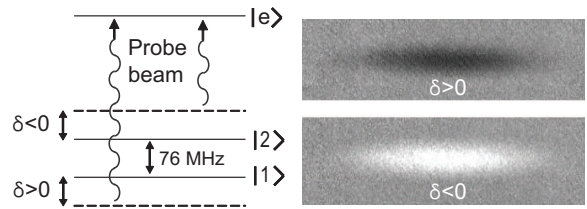}
\caption{Phase-contrast imaging of a balanced spin mixture in
states $\ket{1}$ and $\ket{2}$. White images (phase shift
$\phi>$0) were obtained for a probe beam red-detuned from the
$\ket{2}\rightarrow\ket{e}$ transition, corresponding to $\delta<0$. Black images
($\phi<$0) were obtained for a probe beam blue-detuned from the
$\ket{1}\rightarrow\ket{e}$ transition, corresponding to $\delta<0$.
\label{imaging}}
\end{figure}

\begin{figure}[!htbp]
\includegraphics[width=3.3in]{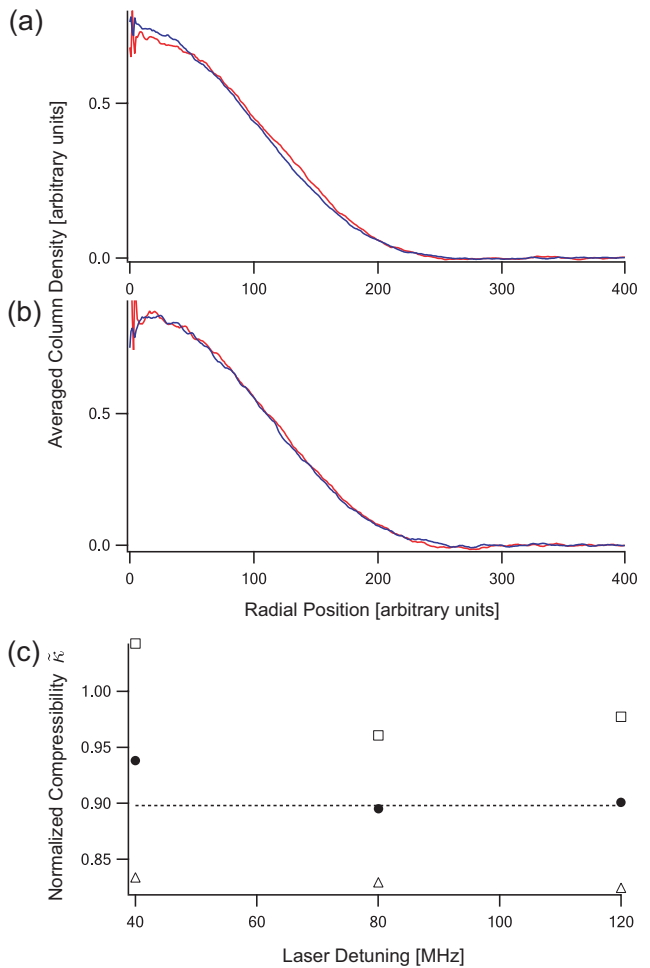}
\caption{Comparison of column density profiles obtained from
positive and negative detuning of (a) 40 and (b) 80 MHz,
respectively. (c) The measured compressibility (at $k_F a=0$) from
positive (open triangle) and negative (open square) detuned
profiles and their averages (solid circle) are shown. The average
value stays constant above 80 MHz detuning. \label{pc}}
\end{figure}

We investigated positive and negative detuning ($\delta$) of 40, 80, and 120 MHz (see Fig.~\ref{imaging}).  The normalized phase
contrast image had a maximum signal of 0.35/1.85, 0.55/1.6,  and 0.7/1.4 for the three positive/negative detuning.  The lensing effect
is opposite (focusing vs. defocusing) for positive and negative detuning, and can therefore be identified by comparing profiles
obtained with positive and negative detuning.  Fig.~\ref{pc} shows that at 40 MHz, the two profiles show a visible difference, but for
profiles at 80 MHz and 120 MHz, the differences are smaller than the noise level.

However, the compressibility is determined by the slope of the profiles and very sensitive to small distortions even if they are
not perceptible in the profiles.  Fig.~\ref{pc}(c) shows that even
at 120 MHz detuning, the compressibilities obtained from profiles
with the two signs of the detuning differ by about 10 \%.  Since
further detuning would have resulted in a smaller signal we
evaluated the average value of the compressibility for positive
and negative.  When the dispersive distortions are small, the
effect on the compressibility should be a first-order effect in
the phase shift and cancel for the average.  Indeed, the average
value stays constant above 80MHz detuning.Our conclusion is that
for reasonable signal levels (i.e. 50 \% of the baseline set by
the probe light) dispersive effects are relevant for quantitative
studies, but can be eliminated by performing averages over
positive and negative detunings.

\section{Discussion}

We address now to what extent a small molecular fraction contributes to the observed density profiles. The presence of molecules is
unavoidable since they form during the ramping and equilibration time. At the highest magnetic field used in the experiment, 679 G, the
molecular fraction was determined to be $\sim$ 10 $\%$ (Fig. 1). Ref. \cite{mol_image} reported that molecules at 650G showed
an absorption cross section of about half the value of the atoms for probe light at the atomic resonance. For phase-contrast imaging
with large detuning molecules and atoms should contribute equally to the signal.

We performed simulations to address how the presence of molecules would affect the compressibility measurements. We considered as
possible scenarios (i) that the molecular fraction is constant throughout the cloud, (ii) that the molecular fraction is proportional
to the loss rate ($n^{8/3}$), and (iii) that the molecular fraction is well equilibrated at the same temperature as atoms. The atomic
profile is then the difference of the measured density profile minus the simulated molecular density distribution.  Scenario (iii) is
ruled out since it would result in a rather sharp peak in the density profile which was not observed.  The first two scenarios with a
10 $\%$ molecular fraction resulted in a value for the normalized compressibility which was increased by 3.3 $\%$ and 4.4 $\%$
respectively. This shows that for our largest value of $k_F a$ the presence of molecules starts to become a systematic effect. In
addition to the contribution to the density profiles, molecules can affect the atomic density distribution through atom-molecule
interactions.

Our work shows that the interaction effect on the compressibility at the maximum possible values of $k_F a$ is about 15 \%.  We could
identify this effect only by careful thermometry (to distinguish it from thermal effects) and by correcting small dispersive
distortions of the cloud. It is desirable to study fermions for stronger repulsive interactions where stronger and non-perturbative
effects are predicted.  The maximum possible $k_F a$ value for obtaining equilibrium density profiles is determined by the loss rate
which is proportional to $n^2 a^6 {\rm max}(T,T_F) = (k_F a)^6 n^{2/3}$ \cite{loss_rate_petrov,Ersy2005}.  Therefore, the maximum possible $k_F a$ for a given loss rate is
proportional to $n^{-1/9}$ and stronger interaction effects can be seen at lower density.  This should be accomplished by reducing the
radial confinement and not the axial confinement which determines the equilibration time. However, the weak density dependence will
allow only modest increases in $k_F a$.  A recent experiment used density ten times smaller than ours \cite{BEC2sus} and reported
ramping from ${k_F}^o a = 0$ to ${k_F}^o a = 0.35$ in 500 ms losing only 5 $\%$ atoms. Assuming the loss happened during the last 50
ms, we can roughly estimate a loss rate of $\sim$ 0.001 ms$^{-1}$ at ${k_F}^o a = 0.3$ which is lower than our measurement, consistent with
the lower density.

Longer lifetimes for a given $k_F a$ should be realized using narrow instead of broad Feshbach resonances. For narrow Feshbach
resonances the low-lying molecular state has a dominant closed-channel character.  Therefore, three body recombination of atoms (which
are in the open channel) has a smaller overlap to the molecular state and therefore a reduced loss rate.  Recent experiments using RF
spectroscopy \cite{narrow_grimm, narrow_Ohara} confirm this.  However, for such narrow resonances the zero-range approximation is no
longer valid, the interaction is no longer described by the scattering length alone and becomes (through an effective range parameter)
momentum dependent. As a result, the narrow Feshbach resonances realize a different Hamiltonian.

In conclusion, in this paper we have addressed the question to what extent Fermi gases with strong interactions can be studied by
observing equilibrium density profiles.  The range of sufficiently long metastability to reach equilibrium is limited to values of $k_F
a < 0.25$. In this range, interaction effects are comparable to thermal effects, but we were able to observe how interactions reduce
the compressibility and obtained quantitative agreement with the first-order mean field term.  An observation of the second order
Lee-Huang-Yang correction is in experimental reach.

If experiments can be performed at stronger interactions, a natural extension of our work would be a measurement of the spin
susceptibility using population imbalanced Fermi systems.  This was performed recently for fermions with attractive interactions
\cite{ENS_sus}.  Such measurements could address the possible existence of a ferromagnetic transition in a repulsive Fermi gas
\cite{FM} for which the spin susceptibility would diverge at the phase transition \cite{string2010}.


\begin{acknowledgments}
This work was supported by the NSF and ONR, an AFOSR MURI, and by ARO
grant no. W911NF-07-1-0493 with funds from the DARPA Optical
Lattice Emulator program. Y.-R. Lee acknowledges support from the Samsung Scholarship. T. T. Wang acknowledges support from NSERC. We are thankful to Christophe Salomon, Christian Sanner, Ariel Sommer, Mark Ku, and
Martin Zwierlein for valuable discussions, and Gregory Lau for
experimental assistance.
\end{acknowledgments}

\bibliography{reference}

\end{document}